\documentclass[twocolumn,showpacs,superscriptaddress,aps,prl]{revtex4}

\usepackage{graphicx}% Include figure files
\usepackage{amssymb,amsmath}
\usepackage{dcolumn}% Align table columns on decimal point
\usepackage{bm}% bold math

\newcommand{\beq}{\begin{equation}}
\newcommand{\eeq}{\end{equation}}

\begin{document}

\preprint{$\#$1 zaliznyak}

\title{Breakdown of the N=0 Quantum Hall State in graphene: two insulating regimes }%Force line breaks with \\
%Breakdown of the QHE in the N=0 Landau Level in graphene

 \author{L. Zhang}
 \author{J. Camacho}
 \affiliation{DCMPMS,
% Department of Condensed Matter Physics and Material Science,
 Brookhaven National Laboratory, Upton, NY 11973 USA}
 \author{H.~Cao}
 \author{Y.~P.~Chen}
 \affiliation{Department of Physics, Purdue University, West Lafayette, IN 47907 USA}
% \affiliation{Birck Nanotechnology Center, Purdue University, West Lafayette, IN 47907 USA}
%
 \author{M.~Khodas}
 \affiliation{DCMPMS,
% Department of Condensed Matter Physics and Material Science,
 Brookhaven National Laboratory, Upton, NY 11973 USA}
 \affiliation{Physics Department,
% Department of Condensed Matter Physics and Material Science,
 Brookhaven National Laboratory, Upton, NY 11973 USA}
 \author{D.~E.~Kharzeev}
 \affiliation{Physics Department,
% Department of Condensed Matter Physics and Material Science,
 Brookhaven National Laboratory, Upton, NY 11973 USA}
 \author{A.~M.~Tsvelik}
 \author{T.~Valla}
 \author{I.~A.~Zaliznyak}
 \email[Corresponding author: ]{zaliznyak@bnl.gov}%
%\thanks{}
 \affiliation{DCMPMS,
% Department of Condensed Matter Physics and Material Science,
 Brookhaven National Laboratory, Upton, NY 11973 USA}

\date{\today}% It is always \today, today,

\begin{abstract}

We studied the unusual Quantum Hall Effect (QHE) near the charge neutrality point (CNP) in high-mobility graphene sample for magnetic fields up to 18 T. We observe breakdown of the delocalized QHE transport and strong increase in resistivities $\rho_{xx},|\rho_{xy}|$ with decreasing Landau level filling for $\nu < 2$, where we identify two insulating regimes. For $1 \gtrsim |\nu| \gtrsim 1/2$ we find an exponential increase of $\rho_{xx,xy} \sim e^{a(H-H_c)}$ within the range up to several resistance quanta $R_K$, while the Hall effect gradually disappears, consistent with the Hall insulator (HI) with local transport.
Then, at $\nu \approx 1/2$ a cusp in $\rho_{xx}(H)$ followed by an onset of even faster growth indicates transition to a collective insulator (CI) state. The likely candidate for this state is a pinned Wigner crystal.

%Transverse resistivity $\rho_{xy}$ eventually becomes independent of %the direction of magnetic field
%Hence, we observe two regimes, first a disorder-dominated resistive $n %= 0$ state and then, a genuine bulk insulating state.

\end{abstract}


%\keywords{}
\maketitle


%\section{}

Graphene is a honeycomb monolayer of C atoms, which forms the most common carbon allotrope -- graphite, but has only recently been isolated in the two-dimensional (2D) form \cite{Novoselov_Science2004}. A bipartite honeycomb $sp^2$-bonded lattice gives graphene its unique electronic structure. Two dispersion sheets associated with the electrons belonging to two different sublattices form the filled $\pi$ and the empty $\pi^*$ bands, which meet at two distinct isolated points (valleys $K$ and $K'$), yielding point-like Fermi surfaces. Low-energy electronic states in each valley have a linear 2D conical dispersion, $\varepsilon({\bf p}) = v_F p$, and in addition to 2D momentum have a 2D pseudo-spin quantum number accounting for the two-sublattice structure. Such quasiparticles are formally described by the Dirac equation for chiral massless fermions and have peculiar transport properties \cite{Beenakker_RMP2008,CastroNeto_RMP2009}.

\begin{figure}[!t]
\begin{center}
\includegraphics[width=.8\columnwidth]{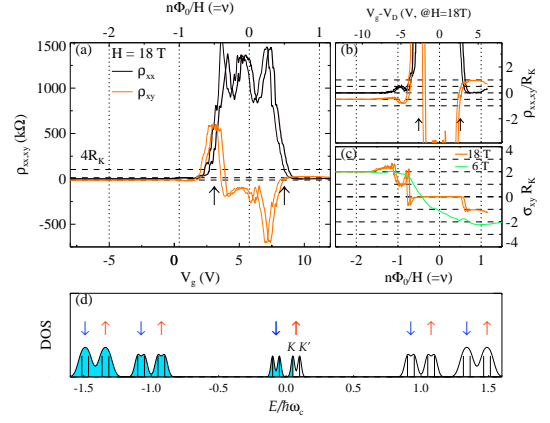}
\caption{Breakdown of the Quantum Hall state in graphene at $H = 18$ T as a function of gate voltage (carrier concentration) at T = 0.25 K. (a) Onset of a strong increase of $\rho_{xx}$ and $|\rho_{xy}|$ beyond $\approx 4 R_K$ near filling $\nu = \pm 1/2$ (shown by arrows). Blow-up of $\rho_{xy}$ in (b) and $\sigma_{xy}$ in (c) show a well-developed QHE plateau around $\nu = -2$ and a hint of plateaux developing at $|\nu| = 1$ at $R_K$ and $1/R_K$, respectively. The apparent $\nu = 0$ plateau in (c) corresponds to a bulk insulator with zero Hall angle and no Hall effect. (d)  Schematics of Landau levels in graphene, including Zeeman splitting (up/down arrows) and $K,K'$ valley splitting (bars).
}
\label{f:QHE-3}
\vspace{-0.25in}
\end{center}
\end{figure}

In a magnetic field $H$ perpendicular to the graphene layer the spectrum of Dirac quasiparticles is quantized into Landau levels (LL) with energies $E_N = \pm \hbar \omega_c \sqrt{N}$. Plus and minus signs correspond to electrons and holes, respectively, $\omega_c = v_F \sqrt{2 e H/(\hbar c)}$ is the ``cyclotron frequency'' for Dirac fermions, and $N = n' + 1/2 \pm 1/2$ where $n' = 0, 1, 2, ...$ enumerates orbital wave functions and $\pm 1/2$ are pseudo-spin eigenvalues. For each $E_N$ there are four states, corresponding to different spin and valley indices. In the presence of Zeeman interaction and inter-valley scattering, these states might split further, as illustrated in Fig.~\ref{f:QHE-3} (d). Unlike the case of Landau quantization for non-relativistic massive electrons, where $E_N =\hbar \omega_c (N + 1/2)$ and $\omega_c = e H/(m c)$, in graphene there is a field-independent level at $E = 0$ for $N = 0$. At the CNP (in undoped graphene) this level is half-filled, being equally shared between particles and holes. Experimentally, such peculiar LL structure is manifested in the unusual QHE observed in graphene, \cite{Novoselov_Nature2005,Zhang_Nature2005}, where Hall conductivity is quantized as $\sigma_{xy} = 4(l + 1/2)/R_K$, $l$ is an integer and $R_K = h/e^2$ is the resistance quantum. The nature of electronic states on the $N = 0$ level remains unclear and has recently become the focus of considerable attention \cite{CastroNeto_RMP2009,Abanin_PRL2007,Giesbers_PRL2007,Jiang_PRL2007,Zhang_PRL2006,Checkelsky_PRL2008}.

Here we report an experimental investigation of charge transport in a high-mobility graphene sample for low carrier density $n$ near the CNP and in magnetic fields up to 18 T, that is, in the QHE regime near the $N=0$ Landau level, where previous studies have yielded conflicting results \cite{Abanin_PRL2007,Giesbers_PRL2007,Jiang_PRL2007,Zhang_PRL2006,Checkelsky_PRL2008}. In particular, some studies have found finite longitudinal resistance $\rho_{xx} \sim R_K$ near the CNP even for high magnetic fields, where the plateau at $\rho_{xy} = R_K/2$ around $\nu = n \Phi_0 / H = \pm 2$  ($\Phi_0$ is the flux quantum) corresponding to either particle or hole filling of the two lowest $N = 0$ LL is well developed \cite{Abanin_PRL2007,Giesbers_PRL2007,Jiang_PRL2007}. Others, reported $\rho_{xx} \gg R_K$, in the M$\Omega$ range, indicating an insulating $N = 0$ state at high fields \cite{Zhang_PRL2006,Checkelsky_PRL2008}. In Ref. \onlinecite{Checkelsky_PRL2008} the $\rho_{xx}(H)$ divergence with $H$ was analyzed as an ad-hoc Kosterlitz-Thouless transition and associated with a critical field $H_c$ (rather than a filling $\nu_c$), which was found to be sample-dependent. The discrepancies between different measurements could be somewhat reconciled by the fact that cleaner samples show stronger divergence of $\rho_{xx}(H)$ with increasing $H$ in the $N = 0$ state \cite{Checkelsky_PRL2008}. Hence, sample quality appears crucial for understanding the physics of the $N = 0$ LL state in graphene.

\begin{figure}[!h!t]
\begin{center}
\includegraphics[width=.8\linewidth]{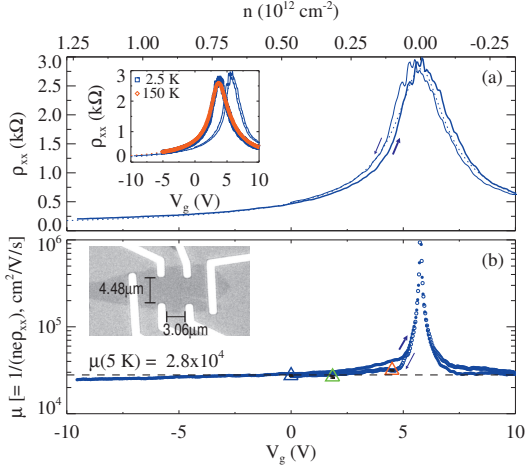}
\caption{Longitudinal resistivity $\rho_{xx}$ of our sample in zero magnetic field as a function of the gate voltage $V_g$. (a) Solid lines are $\rho_{xx}$ at T = 5 K for increasing (thick) and decreasing (thin) $V_g$, as indicated by arrows. Small hysteresis results from mobile charges in Si/SiO$_2$ substrate. The inset shows the same data together with the results of similar follow-up measurements at T = 2.5 K and T=150 K. Dotted lines are fits to $\rho_{xx} = 1/ \sqrt{\sigma_{min}^2 + (e \mu n)^2}$. (b) Drude mobility, $\mu = 1/(ne\rho_{xx})$, \cite{Tan_etal_Kim_PRL2007}. The dashed line shows $\mu = 2.82(6) \cdot 10^4$ cm$^2$/V/s resulting from the fit. Filled and open symbols correspond to different $V_g$ sweep directions shown by arrows. Triangles are Hall mobilities obtained from fits to the data in Fig.~~\ref{f:QHE-1}(d). The inset is the optical image of our sample.}
\label{f:FET-mob}
\vspace{-0.25in}
\end{center}
\end{figure}
We have studied a monolayer graphene sample prepared by mechanical exfoliation of ZYA grade highly oriented pyrolytic graphite (HOPG) on a Si/SiO$_2$ substrate using the standard procedure described in \cite{Novoselov_Science2004}. Transport properties summarized in Fig.~\ref{f:FET-mob} (a,b) reveal high quality of our sample. Magnetoresistance measurements were performed using the low frequency ($17.777$ Hz) lock-in technique with a driving current $I = 10$ nA and the 18 T superconducting magnet at the National High Magnetic Field Laboratory (NHMFL). The follow-up measurements in fields up to 6 T were performed at Purdue University. The sample mobility between the two measurements remained unchanged within the error, $\mu = 2.82(6) \cdot 10^4$~cm$^2$/V/s, while the CNP has shifted by $\approx 2$~V. At T = 150 K, $\mu$ decreased by less than 10\%, to $\mu (150 K) \approx 2.6 \cdot 10^4$~cm$^2$/V/s. A slight electron-hole asymmetry of the resistance in Fig.~\ref{f:FET-mob} is typical of devices with invasive contacts (see inset in Fig.~\ref{f:FET-mob}, (b)) and is explained by the work function difference between graphene and the Au/Cr electrodes in our device \cite{Huard_PRB2008}.

\begin{figure}[!t]
\begin{center}
\includegraphics[width=1.\columnwidth]{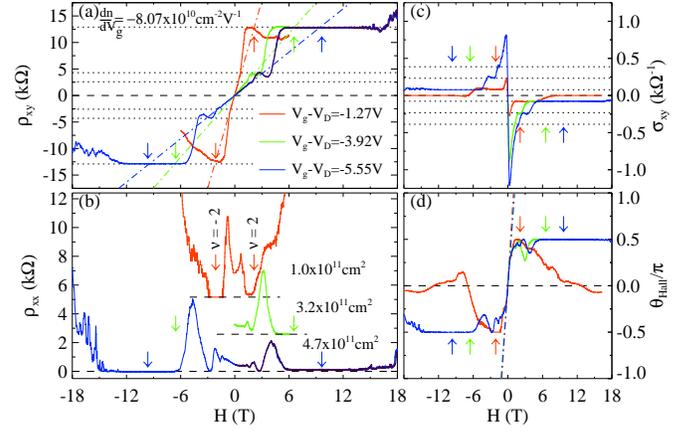}
\caption{QHE in our sample for different gate offsets $V_g - V_D$ from the Dirac point $V_D$, i. e. for different carrier densities $n$. (a) $\rho_{xy}$, (b) $\rho_{xx}$, (c) $\sigma_{xy}$, and (d) Hall angle, $\theta_\mathrm{Hall}$. $\nu = \pm 2$ filling for each $n$ is shown by the corresponding arrow. Dash-dot lines in (a) are linear fits of the low-field part of $\rho_{xy}(H)$ used to refine the Hall constant and obtain $n = H/(e c \rho_{xy}(H))$. Curves in (b) are shifted for clarity, the zero level for each is given by the broken horizontal line. Horizontal dotted lines in (a) and (c) show the resistance and the conductivity quantization in graphene. }
\label{f:QHE-1}
\vspace{-0.25in}
\end{center}
\end{figure}
Fig.~\ref{f:QHE-1} summarizes the QHE in our sample for three charge carrier densities near the CNP, $n = 1 \cdot 10^{11}$, $ 3.2 \cdot 10^{11}$ and $ 4.7 \cdot 10^{11}$~cm$^{-2}$. The latter were refined by fitting the low-field linear part of the Hall resistivity to $\rho_{xy}(H) = H/ (n e c) $, Fig.~\ref{f:QHE-1} (a). This yields $dn/dV_g = -8.07(9) \cdot 10^{10}$ cm$^{-2}$/V. Clear plateaux corresponding to half the resistance quantum in $\rho_{xy}$ and zero $\rho_{xx}$ develop for all three carrier densities upon approaching the LL filling $|\nu| = 2$, Fig.~\ref{f:QHE-1} (a) -- (c). This is a hallmark of the QHE in graphene, resulting from only two of the total four $N = 0$ LL being available to either electrons, or holes, Fig.~\ref{f:QHE-3} (d). Hence, two conducting channels and $\rho_{xy} = \pm R_K/2$ plateaux. The developed QHE regime is also manifested by the plateau at $\pi/2$ in the Hall angle, $\theta_\mathrm{Hall} = \mathrm{atan}(\rho_{xy}/\rho_{xx})$, Fig.~\ref{f:QHE-1} (d). Fitting the low-field linear part of $\theta_\mathrm{Hall}$ to the result of the Boltzman transport theory, $\mathrm{tan}(\theta_\mathrm{Hall}) = (\mu/c) H$, allows for an alternative refinement of the sample mobility $\mu$ (dash-dotted lines in Fig.~\ref{f:QHE-1} (d)). Thus obtained Hall mobility agrees very well with the Drude field mobility, Fig.~\ref{f:FET-mob} (b).

\begin{figure}[!t]
\begin{center}
\includegraphics[width=1.\linewidth]{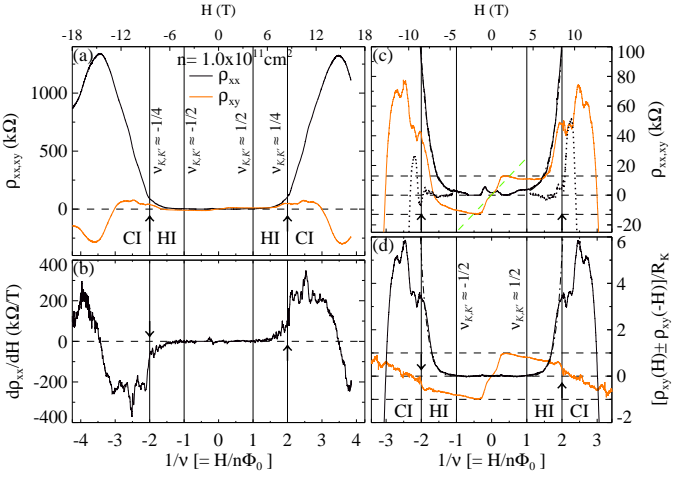}
\caption{Breakdown of QHE in graphene for $n \approx 10^{11}$ cm$^{-2}$ and T = 0.25 K. Top scale shows magnetic field $H$ corresponding to the inverse LL filling $1/\nu$ on the bottom. Filling per valley is $\nu_{K,K'} \approx \nu/2$. (a) $\rho_{xx}(H)$ (dark) and $\rho_{xy}(H)$ (light); (b) $d\rho_{xx}(H)/dH$. Dramatic increase in $\rho_{xx}(H)$ beyond a cusp at $\approx 4 R_K$ and a step-like anomaly in $d\rho_{xx}(H)/dH$ indicate transition to a collective insulator state at $\nu \approx \pm 1/2$. (c) blow-up of the initial growth in $\rho_{xx,xy}(H)$ and (d) the sum (dark) and the difference (light) of $\rho_{xy}$ for two opposite directions of magnetic field show disappearance of the Hall component following breakdown of $|\nu| = 2$ QHE state with increasing $H$ (see also Fig.~\ref{f:QHE-1}(a,b)). Dash-dotted curves in (c), (d) nearly coincident with the data in the HI phase at $1 \gtrsim |\nu| \gtrsim 1/2$ are fits to an exponential increase, $\rho_a(H) = a e^{b(|H|-H_c)}$ \cite{Giesbers_etal_arXiv2009}. Dotted curve in (c) is the difference, $\rho_{xx}(H) - \rho_a(H)$.}
\label{f:QHE-2}
\vspace{-0.25in}
\end{center}
\end{figure}

When LL filling decreases far enough past $|\nu| = 2$ with the increasing magnetic field, the QHE resistance quantization breaks down and the system enters a resistive state, where $\rho_{xx}(H) > 0$ and fluctuates widely with $H$. This fluctuating behavior of $\rho_{xx}(H)$ and $\rho_{xy}(H)$ in Fig.~\ref{f:QHE-1} (a), (b) is reproducible. In fact, the $H > 0$ parts of the curves for $n = 4.7 \cdot 10^{11}$~cm$^{-2}$ overlay results of three different field sweep measurements, which essentially coincide. While $\nu \gtrsim 1$ for this carrier density in our field range and $\rho_{xx,xy}(H)$ increase only moderately, much lower $N = 0$ LL fillings, $\nu \ll 1$, are achievable for $n = 10^{11}$~cm$^{-2}$. Here we observe dramatic increase of $\rho_{xx,xy}(H)$, which is detailed in Fig.~\ref{f:QHE-2}. In terms of the Hall conductivity it appears as a ``zero plateau'' state with $\sigma_{xy} = 0$ around $\nu = 0$,  Fig.~\ref{f:QHE-1} (c). It is important to notice, however, that the Onsager symmetry $\rho_{xy}(H) = -\rho_{xy}(-H)$ is violated in this state, indicating breakdown of the dissipation-less QHE transport via delocalized states,
%not only a but also of an equilibrium transport in general,
typical of a Hall insulator (HI) \cite{Shahar_Nature1998}. This behavior is also reflected by the Hall angle dependence in (d), which deviates from the $|\theta_\mathrm{Hall}(H)| = \pi/2$ QHE plateau and tends towards zero in high magnetic field.

Fig.~\ref{f:QHE-2} shows the full range of $\rho_{xx}(H)$ and $\rho_{xy}(H)$ variation for $n = 10^{11}$~cm$^{-2}$, whose initial parts are also included in Fig.~\ref{f:QHE-1}. Careful examination of the data allows identifying distinct transport regimes. First, for some $\nu$ in the range $1 \lesssim \nu \lesssim 2$ the QHE resistance quantization breaks down and a resistive state forms. The Hall angle becomes $|\theta_\mathrm{Hall}| < \pi/2$, but the Onsager relation still holds, as shown by the sum of Hall resistivities for two opposite magnetic field directions, $\rho_{xy}(H) + \rho_{xy}(-H) \approx 0$ in Fig.~\ref{f:QHE-2} (d). Such behavior could be a sign of the full splitting of $N = 0$ level shown in Fig.~\ref{f:QHE-3} (d) and of a developing $\nu = 1$ plateau, or of the unusual resistive Hall metal (HM) state at $N = 0$ LL considered in \cite{Abanin_PRL2007,Shimshoni_2008}.

In the $1/2 \lesssim \nu \lesssim 1$ regime, a field-symmetric component breaking the Onsager relation appears in transverse resistivity $\rho_{xy}$. The Hall component $\rho_{xy}(H) + \rho_{xy}(-H)$ decreases, while $\rho_{xx}(H)$ and $\rho_{xy}(H)$ steadily increase up to several resistance quanta. Both $\rho_{xx}(H)$ and the symmetric part, $\rho_{xy}(H) +\rho_{xy}(-H)$, follow the same exponential dependence, $\rho_a(H) = a e^{b(|H|-H_c)}$, shown by dash-dot lines in Fig.~\ref{f:QHE-3} (c,d). Independent fits in panels (c) and (d) give consistent $H_c = 4.6(2)$ T, corresponding to $\nu \approx 1$, or a filling $\nu_{K,K'} \approx \nu/2 \approx 0.5$ per valley.

Such behavior can be understood as a Hall insulator (HI) \cite{Abanin_PRL2007,Shahar_Nature1998} resulting from the Zeeman splitting of the $N=0$ LL, Fig.~\ref{f:QHE-3} (d). In this case the delocalized Quantum Hall state and the related $N =0$ mobility edge shift to a finite energy $E_0 \approx g \mu_B H$ (g is the Lande factor and $\mu_B$ is the Bohr's magneton). For filling factors below $\nu \approx 1$ the chemical potential falls below $E_0$ and the electrons are localized. Thermally activated Hall transport via delocalized states is possible for $T > 0$, but it vanishes with the increasing field-induced spin splitting of the $N = 0$ level.
%but is in the regime of the electrical breakdown imposed by the %driving current, $I = 10$ nA in our measurement.
At $T \approx 0$ the system is insulating, the transport occurs via hopping between localized states and is dominated by the mesoscopic conductance fluctuations in the sample. The transverse voltage drop results from the sample and the leads average asymmetry. The $I-V$ curve is expected to be strongly non-linear, as it was indeed observed in Ref. \onlinecite{Checkelsky_PRL2008}, although authors there have interpreted this nonlinearity as resulting from sample heating.
%When this paper was finalized for submission
A similar $N = 0$ insulating regime in a lower mobility sample was recently reported in Ref. \cite{Giesbers_etal_arXiv2009}.

Probably the most surprising and intriguing is the $\nu \lesssim 1/2$ regime, which is characterized by dramatic changes in behavior of $\rho_{xx}(H)$ and $\rho_{xy}(H)$. Both $\rho_{xx}(H)$ and the symmetrized $\rho_{xy}(H) +\rho_{xy}(-H)$ deviate from the exponential growth describing the HI phase. The difference $\rho_{xx}(H) - \rho_a(H)$ shown by the dotted line in Fig.~\ref{f:QHE-2} (c) emphasizes a cusp in $\rho_{xx}(H)$, which is also visible in Fig.~\ref{f:QHE-2} (a), and is followed by an abrupt increase of resistance beyond $\rho_{xx} \sim 4R_K$. The cusp-like singularity in $\rho_{xx}(H)$ is also identified by the prominent jump in the derivative $d\rho_{xx} /dH$ at $\nu \approx 1/2$, Fig.~\ref{f:QHE-2} (b). The Hall component of the transverse resistivity given by the difference $\rho_{xy}(H) - \rho_{xy}(-H)$ also shows singular behavior, abruptly approaching zero at the same filling, Fig.~\ref{f:QHE-2} (d), so that $\rho_{xy}(H) \approx \rho_{xy}(-H)$ for $\nu \lesssim 1/2$.

Simultaneous singular behavior of $\rho_{xy}(H) - \rho_{xy}(-H)$ and $d\rho_{xx} /dH$ at $\nu \approx 1/2$ indicate transition to a different, more insulating state, which we identify as collective insulator (CI). A likely candidate for such a CI state is a pinned Wigner crystal (WC), where electrons are collectively, rather than individually localized.
Our interpretation is based upon analogy with QHE in high-mobility 2D electron gases (2DEG) in semiconductor heterostructures, where an onset of strongly insulating behavior at low filling factors, $\nu \lesssim 1/4$, has also been reported \cite{Jiang_PRL1990,Suchalkin_JETPLett2001,YPChen_NatPhys2006}. This collective-insulating behavior in 2DEG is understood as a 2D WC that can be pinned by arbitrarily small disorder at zero T \cite{Merkt_PRL1996,Fertig_PRB1999}. In our case of graphene transition at $\nu \approx 1/2$ corresponds to a filling $\nu_{K,K'} \approx 1/4$ per valley, roughly consistent with 2DEG results.

Finally, we have also investigated the breakdown of the $N = 0$ Quantum Hall state and the appearance of the insulating behavior in our graphene sample by varying the gate voltage $V_g$ in magnetic field $H = 18$ T. The results shown in Fig. \ref{f:QHE-3} are in agreement with those in Fig. \ref{f:QHE-2}, supporting the picture presented above. There is a well-developed QHE plateau around filling $|\nu| = 2$. The quantization breaks down in the $1 \lesssim |\nu| \lesssim 2$ range and there is a hint of a plateau at $|\rho_{xy}| = R_K$ developing around $|\nu| \approx 1$, Fig. \ref{f:QHE-3} (b), (c). An onset of strong increase of $\rho_{xx}$ and $|\rho_{xy}|$ beyond $\sim R_K$ is seen at $|\nu| \approx 1/2$. It can also be traced in the behavior of the derivative $d\rho_{xy} /dV_g$. The fact that such an abrupt onset occurs here and in Fig.~\ref{f:QHE-2} at very different $|n|$ and $|B|$ but similar $|\nu|=|nh/eH| \approx 1/2$ , suggests that it is driven by the correlation between electrons in graphene, again consistent with a transition to a bulk collective insulator at $|\nu| \lesssim 1/2$. Due to the contact-induced electron-hole asymmetry, there is a slight shift of the curves in Fig. \ref{f:QHE-3} with respect to the nominal $n = 0$, making such measurements by sweeping $V_g$ less favorable compared to the sweep of magnetic field at a constant $V_g$.

In summary, we have investigated the breakdown of the Quantum Hall Effect and the emergence of an insulating behavior in the $N = 0$ Landau Level in a high-mobility single-layer graphene sample. The LL filling in the range $|\nu| \lesssim 2$ is achieved by either increasing the magnetic field at a constant carrier density $n$, or by varying $n(V_g)$ at $H = 18$ T. Careful analysis of our data lead us to identify two different resistive regimes as a function of the decreasing LL filling $\nu = n \Phi_0/H$. First, the well-developed resistance quantization on the $|\nu| = 2$ plateau breaks down and a dissipative state develops near the LL filling $|\nu| \approx 1$. This can be understood as a HI resulting from the Zeeman splitting of the $N = 0$ level, similar to that observed for higher $N$ states \cite{Giesbers_PRL2007,Zhang_PRL2006,Giesbers_etal_arXiv2009}. For $N = 0$, however, as a result there remain no delocalized states occupied by electrons (holes) at sufficiently low filling $|\nu| \lesssim 1$ Fig. \ref{f:FET-mob} (a), and the transport is local at $T \rightarrow 0$. This observation agrees with recent findings reported in Ref. \onlinecite{Giesbers_etal_arXiv2009} and largely rules out the picture of a Hall metal with spin-polarized chiral currents \cite{Abanin_PRL2007}, although not its refined version in \cite{Shimshoni_2008}, which effectively leads to a HI.

Our most striking finding, though, is a well-defined onset of the dramatic resistance increase with decreasing filling at $\nu \approx 1/2$. It is clearly defined by the anomalies in the field dependencies  $\rho_{xx}(H)$ and $\rho_{xy}(H)$ and can be understood as a transition to the collective bulk insulator state. Collective bulk insulating states have been previously observed in 2DEG systems at low filling $\nu$ and are commonly associated with pinned Wigner crystals \cite{Jiang_PRL1990,Suchalkin_JETPLett2001,YPChen_NatPhys2006,Merkt_PRL1996,Fertig_PRB1999}. We suggest that a pinned WC is also likely candidate for the strongly insulating state which we have identified in our graphene sample for $|\nu| \lesssim 1/2$. %($|\nu_{K,K'}| \approx 1/4$)

We thank I. Childres, J.-H. Park and E. Palm for help with the measurements, and S. Suchalkin, Z. Jiang, M. Strongin, D. Abanin and A. Shytov for discussions. This work was supported by the US DOE under the Contract DE-\-AC02-\-98CH10886. Partial support by Purdue University and Miller Family Endowment is gratefully acknowledged. Work at the NHMFL is also supported by NSF through DMR-0084173 and by the State of Florida.

%\newpage %Just because of unusual number of tables stacked at end
%\bibliographystyle{apsrev}
%\bibliographystyle{prsty}

\end{document}